# High temporal resolution total-body dynamic PET imaging based on pixel-level time-activity curve correction

Zixiang Chen[1,2#], Yaping Wu[3#], Na Zhang[1,2], Yu Shen[3], Hairong Zheng[1,2], Dong Liang[1,2], Meiyun Wang[3] and Zhanli Hu[1,2*]

[1]Lauterbur Research Center for Biomedical Imaging, Shenzhen Institute of advanced technology, Chinese academy of science, Shenzhen 518055, China.
[2]Chinese Academy of Sciences Key Laboratory of Health Informatics, Shenzhen 518055, China.
[3]Department of Medical Imaging, Henan Provincial People's Hospital & People's Hospital of Zhengzhou University, Zhengzhou 450003, China.
[4]Central Research Institute, Shanghai United Imaging Healthcare, Shanghai 201807, China.
#These authors contributed equally.
*Author to whom any correspondence should be addressed. (Email: zl.hu@siat.ac.cn)

## Abstract

Dynamic positron emission tomography (dPET) is currently a widely used medical imaging technique for the clinical diagnosis, staging and therapy guidance of all kinds of human cancers. Higher temporal imaging resolution for the early stage of radiotracer metabolism is desired; however, in this case, the reconstructed images with short frame durations always suffer from a limited image signal-to-noise ratio (SNR) and unsatisfactory image spatial resolution. The appearance of uEXPLORER (United Imaging Healthcare, Inc.) with higher PET imaging sensitivity and resolution may help solving this problem. In this work, based on dynamic PET data acquired by uEXPLORER, we proposed a dPET processing method that denoises images with short frame durations via pixel-level time-activity curve (TAC) correction based on third-order Hermite interpolation (Pitch-In). The proposed method was validated and compared to several state-of-the-art methods to demonstrate its superior performance in terms of high temporal resolution dPET image noise reduction and imaging contrast. Higher stability and feasibility of the proposed Pitch-In method for future clinical application with high temporal resolution (HTR) dPET imaging can be expected.

# I. Introduction

Dynamic positron emission tomography (dPET) imaging with $^{18}$F-fluorodeoxyglucose ($^{18}$F-FDG) as the radiotracer is currently a widely used imaging technique to assist in clinical oncological diagnosis [1-4]. Compared to traditional static positron emission tomography (PET) examination, dPET images provide temporal evolution information on imaged tissues, and parametric images revealing the metabolic rate, which is obtained via appropriate image post-analysis, can provide oncologists with extra clinical assistance [5, 6]. Higher imaging temporal resolution, namely, reconstructing images from raw data with shorter frame durations, is desired for dPET imaging, especially for early frames that depict the initial metabolism of the radiotracer. However, the shorter the frame durations are, the more serious the image noise is in the corresponding reconstructed images, and this technical bottleneck of the mutual constraint between spatial and temporal resolutions attracts much attention from scholars working on PET imaging techniques. The core of high temporal resolution (HTR) dPET imaging is to improve the imaging quality with short frame durations. A landmark in modern PET imaging systems for breaking the spatiotemporal resolution bottleneck is the coming out of the first single-bed total-body PET/computed tomography (CT) imaging system, uEXPLORER, developed by United Imaging Healthcare Inc., which is expected to have the best PET imaging quality based on the longest axial field-of-view (FOV) of approximately 2 meters[7].

In addition to the development of clinical imaging system hardware, research on advanced imaging algorithms also promotes the realization of HTR dPET imaging. Many advanced dPET image reconstruction algorithms have been proposed to find better solutions for dPET reconstructions. Early academic progress in dPET image reconstruction includes interpolation basis functions, patient-specific motion models, wavelet denoising, and interframe filtering in mathematical models for image reconstruction [8-14]. A landmark of reconstruction algorithm development was the proposal of the kernel method by Wang et al. in 2015, in which prior dynamic information was creatively applied to construct a kernel matrix that was assembled in the imaging model [15]. Several trials for the further application of prior or supplementary information for HTR dPET image reconstruction, including anatomical information from computational tomography (CT) images or magnetic resonance (MR) images and spatiotemporal prior information based on the kernel method, were proposed [16-18]. Furthermore, classical image denoising techniques, such as nonlocal mean (NLM) and highly constrained back-projection (HYPR), were combined into the kernel method reconstruction [19-21]. In addition to these kernel-based algorithms, Zhang et al. proposed an iterative reconstruction method that accounted for both low-rank matrix constraints and nonlocal total variation (NLTV) regularization and achieved satisfying imaging results [22].

Learning-based methods for denoising dPET images are currently another valuable research field for imaging quality development. Many proposed learning-based methods have been proven to have the ability to address the challenge of HTR dPET imaging. Several deep-learning networks based on a U-net framework work as reported, which use static PET images as network inputs and dynamic images as target outputs [23-25]. Another method based on a convolutional neural network (CNN) that requires the supplementary MR image as the co-input was proposed by He et al. for effective dPET image denoising [26]. Early neural network algorithms, such as the deep auto-encoder (DAE) and multi-layer perceptron (MLP), can be applied to denoise time-activity curves (TACs) of dPET image data [27, 28]. Learning-based methods do make sense; however, one of their remaining issues, from the perspective of industrial application, is model generalization with multicenter data, even for models that require no high-quality training labels.

According to the state-of-the-art studies reviewed above, in terms of advanced algorithms for dPET imaging, regardless of image reconstruction or image posttreatment, it is significant to consider the characteristics of the dynamic evolution of the radiotracer that is used. In other words, isolated frame-by-frame processing is ineffective, especially for HTR imaging that involves frames with a low signal-to-noise ratio (SNR). The successes achieved by many recently proposed dPET image denoising algorithms based on only the current imaging task can better illustrate this statement. One of the most classical methods for dynamic image denoising, the HYPR method, which was first proposed for dynamic MR image processing [29], was applied to dPET image denoising by Christina et al. in 2010. The HYPR method takes advantage of a composite weighted image averaged from the whole dynamic image series to process dPET images [19]. The NLM method was applied to dPET denoising by Dutta et al. This is, however, more complicated than the conventional NLM theory since spatiotemporal patches are involved [30]. Another NLM-based method was proposed by Jomma et al., which includes wavelet processing, a temporally extended NLM filter and Shearlet transform in the processing framework [31]. Floberg et al. proposed a spatiotemporal expectation-maximization (STEM) filtering method, which takes full advantage of Gaussian image filtering while preserving the image resolution by expectation-maximization (EM)-based iterative deconvolution [32]. Most recently, a kinetics-induced block matching and 5-D transform domain filtering (KIBM5D) method was proposed by Hashimoto et al. in 2020. Many efficient concepts, such as batch matching, spectrum transformation and Wiener filtering, were employed in the KIBM5D method, and remarkable improvement in imaging quality was achieved [33].

The motivation of the current study is that a more powerful denoising algorithm with higher stability and feasibility for clinical application is still needed for achieving HTR imaging. Inspired by the DAE and MLP methods, which are reviewed above, that dPET images can be denoised in the temporal domain for TACs instead of in the spatial domain for image patches, we propose a pixel-level TAC correction method based on third-order Hermite interpolation (Pitch-In) to denoise early frames with very short frame durations, which involves no training stage and is based purely on the current imaging task and data.

The rest of this article is organized as follows: Section II gives a detailed demonstration of the Pitch-In method, and Section III introduces the experiments carried out in this study. The experimental results are given in Section IV. We give some discussion on the proposed method in Section V, and the conclusion of the current study is given in Section VI.

## II. Method

### A. Summary of the proposed method

The proposed dPET image denoising method is constructed based on the theory of third-order Hermite interpolation (THI), a classical concept of data processing. This is combined with appropriate strategies for image data fidelity. Data denoising (smoothing) and data fidelity are alternately and iteratively carried out during the image processing procedure. The following sections give detailed illustrations of different modules of the Pitch-In method.

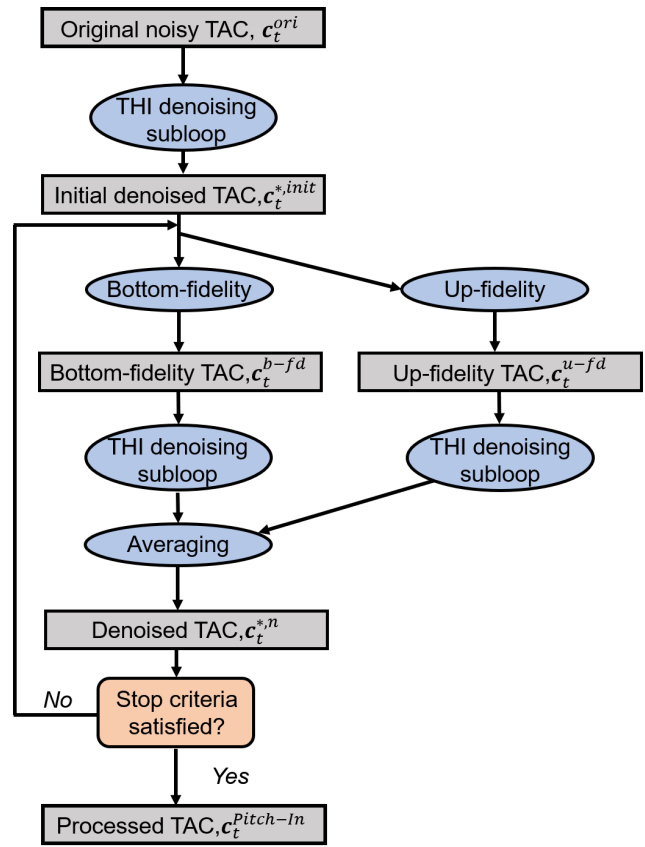

**Figure 1**. Flowchart of the proposed Pitch-In method.

### B. Third-order Hermite Interpolation (THI)

The theory of Hermite interpolation has existed for a long time and has been applied to the development of industrial theories in many fields. In this study, the proposed method takes advantage of a modified Hermite interpolation method proposed by Akima in 1970 [34]. This interpolation method involves a larger range of reference points for the acquisition of one target point; therefore, the estimated curve coincides better with the tendency of the original curve. *Algorithm I* demonstrates how THI is applied in this work for interpolating estimation. This interpolating calculation is

denoted as
$$y_t = THI(x_r, y_r, x_t) \tag{1}$$
where $x_t$ are the coordinates corresponding to values $y_t$ to be estimated, and $x_r$ and $y_r$ are the sampled coordinates and values acting as the reference data for the interpolation.

---

**Algorithm I. Third-order Hermite Interpolation (THI)**

**Function form:** $y_t = THI(x_r, y_r, x_t)$

**Input:** Sampled coordinates, $x_r$; Sampled curve values, $y_r$; Target coordinates, $x_t$

**Output:** Interpolated curve, $y_t$

    *# Curve slope calculation*

    $l \leftarrow length(y_r)$

    **for** $i \leftarrow 1$ **to** $(l-1)$

        $\delta_i \leftarrow \frac{y_{r_{i+1}} - y_{r_i}}{x_{r_{i+1}} - x_{r_i}}$

    **end for**

    $\delta_l = \delta_{l-1}$

    *# Curve derivative calculation*

    **for** $i \leftarrow 3$ **to** $(l-1)$

        $w_1 \leftarrow |\delta_{i+1} - \delta_i| + \frac{|\delta_{i+1} + \delta_i|}{2}$

        $w_2 \leftarrow |\delta_{i-1} - \delta_{i-2}| + \frac{|\delta_{i-1} + \delta_{i-2}|}{2}$

        $y'_i \leftarrow \frac{w_1}{w_1+w_2}\delta_{i-1} + \frac{w_2}{w_1+w_2}\delta_i$

    **end for**

    $y'_1 \leftarrow \delta_1;\ y'_l \leftarrow \delta_l$

    $y'_2 \leftarrow \frac{(2|\delta_2 - \delta_1| + |\delta_2 + \delta_1|)\delta_1 + 2|\delta_1|\delta_2}{2|\delta_1| + 2|\delta_2 - \delta_1| + |\delta_2 + \delta_1|}$

    *# Interpolation*

    $l \leftarrow length(x_t)$

    **for** $i \leftarrow 1$ **to** $l$

        **if** $x_{t_i} \in x_r$

            $y_{t_i} \leftarrow y_r(x_{t_i})$

        **else**

            find neighbors of $x_{t_i}$ in $x_r$, $(x_0, y_0, y'_0)$ and $(x_1, y_1, y'_1)$, where $x_0 < x_{t_i} < x_1$

            $y_{t_i} \leftarrow \left[\left(1 + 2\frac{x_{t_i} - x_0}{x_1 - x_0}\right)y_0 + (x_{t_i} - x_0)y'_0\right]\left(\frac{x_{t_i} - x_1}{x_0 - x_1}\right)^2$

                $+ \left[\left(1 + 2\frac{x_{t_i} - x_1}{x_0 - x_1}\right)y_1 + (x_{t_i} - x_1)y'_1\right]\left(\frac{x_{t_i} - x_0}{x_1 - x_0}\right)^2$

        **end if**

    **end for**

    **return** $y_t$

## C. Pixel-level TAC Smoothing Based on THI

A pixel-level TAC of the dPET image can be described simply by

$$c_t = [c_{t_1}, c_{t_2}, \ldots, c_{t_T}], \quad (2)$$

where $T$ is the total number of effective time frames under image processing. Using THI, we can re-estimate some of the data points in $c$, denoted as the target points, from the rest of the data points, the reference points, in $c$. In this article, the operation of TAC re-estimation based on THI is denoted as

$$c_t^{re-est} = ReTHI(c_t) \quad (3)$$

where $c_t$ is the input TAC to be re-estimated. In this process, we randomly assign $n_t$ data points in $c_t$ to act as references, based on which the remaining points are re-estimated using THI. This operation is illustrated in detail by *Algorithm II*. The TAC denoising step is carried out based on this basic process of data re-estimation: The function *ReTHI*(.) were applied in parallel $k_p$ times for the input TAC of the current denoising epoch, $c_t^{*,I_t-1}$ ($I_t$ denotes the epoch index), and the resultant re-estimated TACs were averaged to obtain $c_t^{update}$, based on which the denoised TAC is given by

$$c_t^{*,I_t} = \eta c_t^{update} + (1-\eta) c_t^{*,I_t-1} \quad (4)$$

where $\eta$ is a relaxation parameter for TAC denoising updating. The subloop of THI denoising consists of $E_t$ denoising epochs, and the whole subloop is denoted as

$$c_t^* = DnTHI(c_t^{in}) \quad (5)$$

In equation (5), the input TAC, $c_t^{in}$, can be either the original noisy TAC or the fidelity TAC (the process of TAC fidelity is described in the next section). Function *DnTHI*(.) can be referred to *Algorithm III*, which gives the details of the THI denoising subloop in our proposed Pitch-In method.

---

**Algorithm II. TAC Re-estimation Based on THI**

---

**Function form:** $c_t^{re-est} = ReTHI(c_t)$

**Input:** Pixel-level TAC, $c_t(t)$

**Parameters:** No. of reference frames, $n_t$; No. of effective frames, $T$

**Output:** Re-estimated TAC, $c_t^{re-est}(t)$

  $I \leftarrow 1\ to\ T$

  $I_r \leftarrow \{Shuffle(I)\}_{1\ to\ n_t}$     *# Reference point random selection*

  $I_t \leftarrow \complement_I I_r$                   *# Objective points of re-estimation*

  $c_{t_i}^{re-est}\big|_{i \in I_t} \leftarrow THI\left(t_i|_{i \in I_t}, t_j|_{j \in I_r}, c_{t_j}|_{j \in I_r}\right)$   *# Algorithm I*

  $c_t^{re-est}(t_i) \leftarrow c_{t_i}^{re-est}$

**return** $c_t^{re-est}(t)$

---

**Algorithm III. THI Denoising Loop**

---

**Function form:** $c_t^* = DnTHI(c_t^{in})$

**Input:** Original noisy TAC or fidelity TAC, denoted as $c_t^{in}$

**Parameters:** No. of parallel *THI* re-estimations, $k_p$; No. of denoising iterations, $E_t$; Relaxation parameter, $\eta$

**Output:** Denoised TAC, $c_t^*$

$c_t^{*,0} \leftarrow c_t^{in}$

for $I_t \leftarrow 1$ to $E_t$

$\quad C \leftarrow$ all-zero matrix of size $k_p \times length(c_t^{in})$

$\quad$ for $p \leftarrow 1$ to $k_p$

$\qquad C(p,:) \leftarrow ReTHI(c_t^{*,I_t-1})$  # Algorithm II

$\quad$ end for

$\quad c_t^{update} \leftarrow \sum_{p=1}^{k_p} C(p,:) / k_p$

$\quad c_t^{*,I_t} \leftarrow \eta c_t^{update} + (1-\eta) c_t^{*,I_t-1}$

end for

$c_t^* \leftarrow c_t^{*,E_t}$

**return** $c_t^*$

## D. TAC Fidelity

Repeated THI denoising steps efficiently help to eliminate the noise in the pixel-level TACs; however, it may also bring concerns about over-smoothing, which makes the concentrations of the radiotracer given by the processed images unreliable without considering data fidelity. A simple method for denoising TAC fidelity based on the original noisy TAC, $c_t^{ori}$, is applied in this study. Given a THI denoised pixel-level TAC, $c_t^*$, we define an up-fidelity group, $I_u$, and a bottom-fidelity group, $I_b$, indicating the corrections of frame indexes where concentrations given by $c^*$ are smaller and larger than those given by $c^{ori}$, respectively. Separated fidelity operations are carried out for these two sets of elements. This fidelity process is denoted as

$$[c_t^{u-fd}, c_t^{u-fd}] = Fidelity(c_t^*, c_t^{ori}). \tag{6}$$

In our article. The up- and bottom-fidelity TACs, $c_t^{u-fd}$ and $c_t^{u-fd}$, respectively, undergo THI denoising processes and are averaged to obtain the input TAC of the next main-loop epoch, $c_t^n$, or the resultant processed TAC, $c_t^{Pitch-In}$. The details of the fidelity process in our processing framework are given in *Algorithm IV*, in which two predetermined relaxation (if >1) or overshoot (if <1) parameters, $\beta_u$ and $\beta_b$ for up- and bottom-fidelity, respectively, as well as the corresponding in-loop-assigned pixel-specific weights, $w^u$ and $w^b$, are involved.

**Algorithm IV. TAC Fidelity**

**Function form:** $\{c_t^{u-fd}, c_t^{u-fd}\} = Fidelity(c_t^*, c_t^{ori})$

**Input:** Denoised TAC, $c_t^*$; Original noisy TAC, $c_t^{ori}$

**Parameters:** Relaxation of overshoot parameters, $\beta_u$ and $\beta_b$

**Output:** Up-fidelity and bottom-fidelity TACs, $c_t^{u-fd}$ and $c_t^{b-fd}$

*# Pixel-specific weights calculation*

$w^u \leftarrow \mathcal{L}_{[0.1\ 1]}\left(exp\left[\mathcal{L}_{[1\ 5]}\left(c_t^* / \sum c_t^*\right)\right]\right)$

$w^b \leftarrow$ all-one vector

*# Find up- and bottom-fidelity groups*

$I_u \leftarrow \{i | c_{t_i}^* \leq c_{t_i}^{ori}\}$

$I_b \leftarrow \{i | c_{t_i}^* > c_{t_i}^{ori}\}$

*# Fidelity*

$c_t^{u-fd}, c_t^{b-fd} \leftarrow c_t^*$

**for** $j$ in $I_u$

$\quad c_{t_j}^{u-fd} \leftarrow c_{t_j}^* + \beta_u w_j^u \left(c_{t_j}^{ori} - c_{t_j}^*\right)$

**end for**

**for** $j$ in $I_b$

$\quad c_{t_j}^{b-fd} \leftarrow c_{t_j}^* - \beta_b w_j^b \left(c_{t_j}^* - c_{t_j}^{ori}\right)$

**end for**

**return** $c_t^{u-fd}\ c_t^{b-fd}$

## E. Processing Strategy and Parameters

The flowchart of the Pitch-In method is given in Figure 1. As shown, the pixel-level TACs alternately undergo THI denoising subloop and fidelity operations to achieve an optimal trade-off between TAC smoothing and TAC fidelity. An initial smoothed TAC, $c_t^{*,init}$, is obtained similarly via a THI denoising subloop with a certain number of epochs from the original noisy TAC before inputting the main loop of the proposed Pitch-In method. One epoch of main-loop data processing consists of a sub-step of data fidelity and a subloop of data THI denoising. The up- and bottom-fidelity TACs, $c_t^{u-fd}$ and $c_t^{u-fd}$, undergo THI denoising subloops to obtain $c_t^{*,u}$ and $c_t^{*,b}$, respectively, and are then averaged to obtain the input TAC, $c_t^n$, for the next processing epoch if the maximum number of epochs has not been reached. The parameters used in this study are listed in Table I, and some discussions on the selection of such parameters are given in Section V.B.

Table I. Parameters of the proposed Pitch-In method.

| Symbol | DESCRIPTION | Value |
|---|---|---|
| $n_t$ | Number of reference points for THI estimations | 0.8×T |
| $k_p$ | Number of parallel estimations in one THI denoising iteration | 20 |
| $E_t$ | Number of iterations of the THI subloop | 30 |
| $E$ | Number of iterations of the main loop | 20 |
| $\eta$ | Relaxation parameter for iterative THI updating | 0.6 |
| $\beta_u$, $\beta_b$ | Relaxation or overshoot parameter for data up-fidelity and bottom-fidelity, respectively | 6 and 1, respectively |

## III. Experiments

In this work, 3D (2D spatial + 1D temporal) simulative dPET image data and clinical total-body dPET data were used in our experiments to validate the proposed Pitch-In method.

### A. Simulative Data

Simulative dPET data were generated from an XCAT thoracic digital phantom. In this phantom, nine types of simulated human tissues were involved, including the aorta, bone, marrow, myocardium, ventricle, lung, liver, tumor and body background. Tissue-specific ground truth TACs representing two time-dividing protocols of the early stage of radiotracer metabolism that were extracted from real total-body dPET image data, as shown in Figure 2a)~b), were assigned to these tissues. The two time-dividing protocols simulated in this work included a low temporal resolution (LTR) imaging protocol yielding 30 frames during one hour of patient scanning and an HTR imaging protocol yielding 60 frames with the same scanning duration, and the detailed frame durations for both LTR and HTR imaging were designed to be the same as that of the employed clinical total-body dPET data. A dynamic scan was conducted with this dynamic phantom with a total simulative photon count of $8.3 \times 10^6$, and simulative images were reconstructed frame by frame based on the acquired sinograms using the traditional expectation-maximization (EM) method [35].

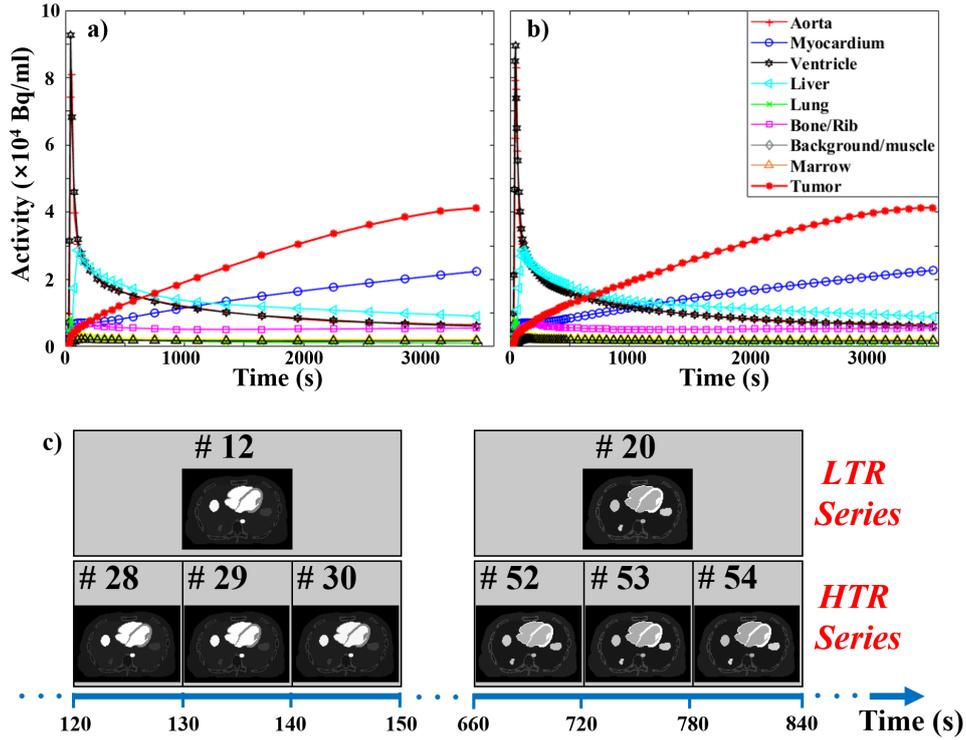

**Figure 2**. Simulated TACs and the applied time dividing protocol. a) Simulated TACs with LTR protocol; b) Simulated TACs with HTR protocol; c) Correspondence relationship between the frames' durations in the HTR and LTR protocols.

## B. Clinical Total-body Image Data

Permitted by the committee of medical ethics of Henan Provincial People's Hospital, total-body dPET images of a patient with lung adenocarcinoma (male, 174 cm height and 88.2 kg weight) were used for the denoising experiment in this work. The dPET image series were collected from the uEXPLORER PET/CT imaging system (United Imaging Healthcare). The dynamic scan lasted for approximately one hour after the injection of $^{18}$F-FDG, and projection data were reconstructed based on two time-dividing protocols: HTR dPET image series including 84 frames (2 s*15, 5 s*6, 10 s*12, 20 s*15, 60 s*12 and 100 s*24) and LTR image series including 30 frames (5 s*6, 10 s*3, 30 s*4, 60 s*5, 180 s*4 and 300 s*8). Images were reconstructed via the 3D time-of-flight (TOF) ordered-subset EM (OS-EM) method [36] with 3 iterations and 20 subsets into a 192*192*673 matrix with a transversal FOV of 600 mm and a slice sickness of 2.886 mm. One thoracic slice with a clearly visible tumor lesion was employed for the experiment in this study.

Framewise duration division was delicately designed in this study. The frame duration of every frame in the LTR series can precisely correspond to two or three frames in the HTR series. For example, as shown in Figure 2c), the frame duration of frame #12 in the LTR series can be equally divided into three parts that correspond to frames #28~#30 in the HTR series, and the manners for other frames are analogous. Note that image processing for both the simulative data and clinical data is carried out for not the total set of dPET data but only the effective frames, which are defined as the frames counting from the peak of the blood input function. More specifically, the simulative and clinical data contained 65 and 67 effective frames, respectively.

## C. Evaluation Methods

For the quantitative evaluation of the proposed Pitch-In method, several quantitative indexes were calculated based on the processed dPET images. The peak signal-to-noise ratio (PSNR) and the structural similarity (SSIM) were calculated frame by frame to evaluate the imaging accuracy in terms of image noise reduction and structure restoration for the simulation experiment. The methods for calculating PSNR and SSIM are the same as those used in our previous work [37]. In addition, boxplot analysis was carried out based on different tissues in several time frames. These tissues include the ventricle, liver and tumor.

The contrast-to-noise ratio (CNR) was used in this study to measure the tissue contrast in the processed images, which is given by:

$$CNR = \frac{\mu_{ROI} - \mu_{BG}}{\sqrt{\sigma_{ROI}^2 + \sigma_{BG}^2}} \tag{7}$$

where $\mu$ and $\sigma$ are the mean activities and standard deviations calculated based on ROIs that are expected to be uniform or background (BG) tissue. In this work, for the simulation experiment, ROIs were extracted from the liver (with size 7*7), ventricle (with size 7*7) and tumor lesion (with size 5*5), and for the clinical data experiment, ROIs were extracted from the aorta (with size 4*4) and the tumor lesion (with size 5*5). The background ROIs were extracted from the regions of the right lung with sizes of 7*7 and 5*5 for the simulative and clinical experiments, respectively.

Patlak Ki parametric imaging was conducted based on the processed dynamic images to evaluate whether the denoising methods skew the dynamic evolution information provided by the originally reconstructed image series. The general Patlak model shows[38]

$$\frac{C(t)}{C_p(t)} = \frac{\int_{t_e}^{t} C_P(\tau)\, d\tau}{C_P(t)} \cdot K_i + b, \tag{8}$$

where $C(t)$ is the measured time-activity curve (TAC) of a voxel, $t_e$ is the reference time point corresponding to the first effective frame, and $C_p(t)$ is the blood input functions extracted from the aorta region of the processed image series. With measured $C(t)$ and $C_p(t)$, Ki is acquired via linear fitting based on Eq. (8). The last 12 and 4 frames, corresponding to the late 20 mins of the HTR and LTR series, respectively, were involved in such linear fitting to estimate the metabolic rate Ki.

## D. Compared Method

Several state-of-the-art dPET image denoising methods were involved for comparison with the proposed Pitch-In method: the HYPR method proposed by Christian et al. in 2010; the STEM method proposed by Floberg et al. in 2013; the spatiotemporal patch-based NLM (NLM-ST) method proposed by Dutta et al. in 2013;

and the KIBM5D methods proposed by Ote et al. in 2020 [19, 30, 32, 33]. Note that for all the methods compared in this work, the implementations require only the data from the current imaging task, and no training stages are needed, which indicates that the clinical applications of these methods can be convenient. The details of the implementation of these compared methods are given as follows.

a) In the **HYPR** method, a Gaussian filter with a size of 23 pixels was used as the spatial filter function, $F$. The size of the temporal sliding window of the composite image, $I_c$, was set to 15 centered at the frame under processing, which helped gather evolution information from the dynamic image series for denoising every frame. The above chosen parameters yield the best performance of the HYPR method [19].

b) In the **STEM** method, a 3D Gaussian filter (2D spatial + 1D temporal) with a spatial full width at half maximum (FWHM) of 4.69 mm and a temporal FWHM of 3 frames, which is set to the same value as that given in Floberg's paper, was applied as the image filter in the STEM method [32]. To achieve satisfying processing results by STEM as much as possible, the number of deconvolution iterations was set to 10.

c) In the **NLM-ST** method, the temporal threshold, $t^*$, was set to the time point corresponding to the first effective frame, which means that the whole set of processed images was included in the temporal component for patch comparison. We included these images as much as possible to avoid involving error tissue structures from the late frames to the early frames. The sizes of the search window and calculation patch were set to 7×7 and 3×3, respectively, and the smoothing parameter $h$ was set to 2.0 to optimize the imaging result of NLM-ST in our work.

d) In the **KIBM5D** method, several parameters were set to the same as in Ote's work, such as the sizes of the spatial neighborhood (4×4 for hard thresholding and 5×5 for Wiener filtering), the sliding step of 3 pixels, the 11×11 search window and the number of nearest grouped patches, which was 32 [33]. The standard deviation $\sigma$ used in the stages of hard thresholding and Wiener filtering was set to 350 in our work to acquire the best performance of KIBM5D that we can access.

## IV. Results

### A. Simulative Experiments

The imaging results of the simulation experiment are shown in Figure 3. HTR images of frames #24, #29, #41 and #50 are shown. The denoising effect of the proposed method is visibly better than that of the HYPR and STEM methods. For NLM-ST and KIBM5D, a remarkable phenomenon is that the background tissue, including the lung and muscle, is smoothed to a greater extent than the hypermetabolic tissues, such as the heart, liver and tumors. In contrast, spatial equilibrium is achieved by the proposed Pitch-In method, of which the denoising effect is not related to tissue intensity. Compared to the other methods, as shown by the marked myocardium tissue, the proposed Pitch-In has the advantage in structure restoration, which may be the result of

Pitch-In processing dynamic images purely in the temporal domain. The quantitative results of PSNR and SSIM values are given in Figure 4. Overall, the proposed Pitch-In method is superior to all the compared methods in terms of image noise reduction. The ability of the structure restoration of the proposed method is further stated by Figure 4d)~e) given the calculated SSIM values.

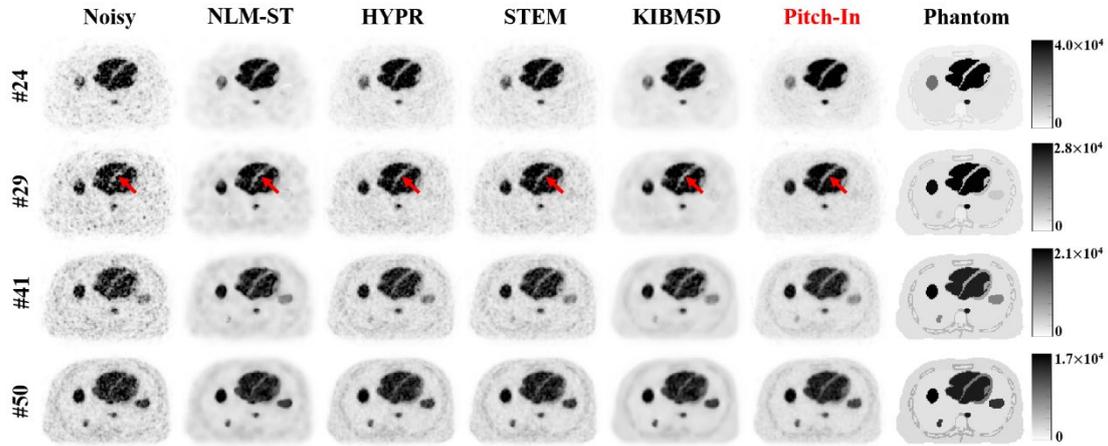

**Figure 3**. Results of the simulative experiment. The unit of the image intensity is Bq/ml, representing the activity concentration of radiotracer in patient tissues.

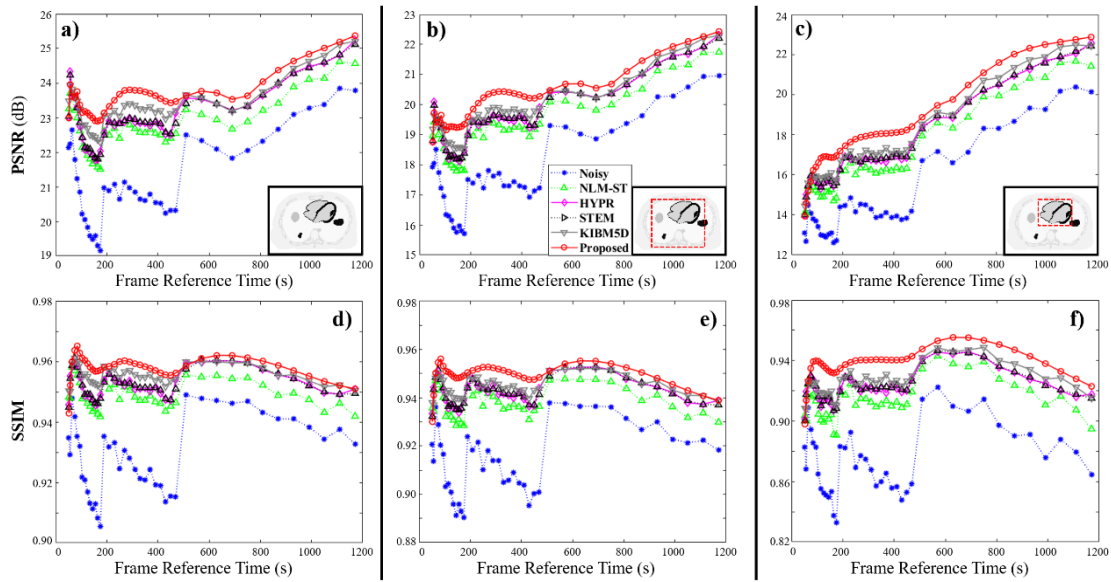

Figure 4. PSNR and SSIM values calculated frame by frame for the simulative images processed by different methods. a) and d) are the PSNR and SSIM values, respectively, calculated based on the whole image space; b) and e) are the PSNR and SSIM values, respectively calculated based on a multi-tissue ROI as indicated by the red dashed box in b); c) and f) are the PSNR and SSIM values, respectively calculated based on the ROI of heart as indicated by the red dashed box in c).

Figure 5 shows the quantitative values of the CNR calculated based on an ROI of different tissues, including the liver, ventricle and a simulative tumor lesion, with respect to an ROI in the background tissue of lung. The Pitch-In method gives the highest CNR values around all the compared methods for most of the frames when considering hypermetabolic tissues such as the ventricle and liver. However, Figure 5c) shows that the proposed Pitch-In method does not yield a higher CNR than the NLM-ST and KIBM5D methods for the tumor lesion, which may be attributed to the over-smoothing of the background region by these two methods. The box plot given in Figure 5d)~f) based on the ROIs in the liver, ventricle and tumor further shows that the proposed method yields the most reliable ROI intensity distribution referring to the true activity concentrations. The superior performance of the proposed Pitch-In method is stated by the achievement of satisfying tissue contrast and effective image noise elimination, while reliable image fidelity and tissue boundary restoration are guaranteed at the same time.

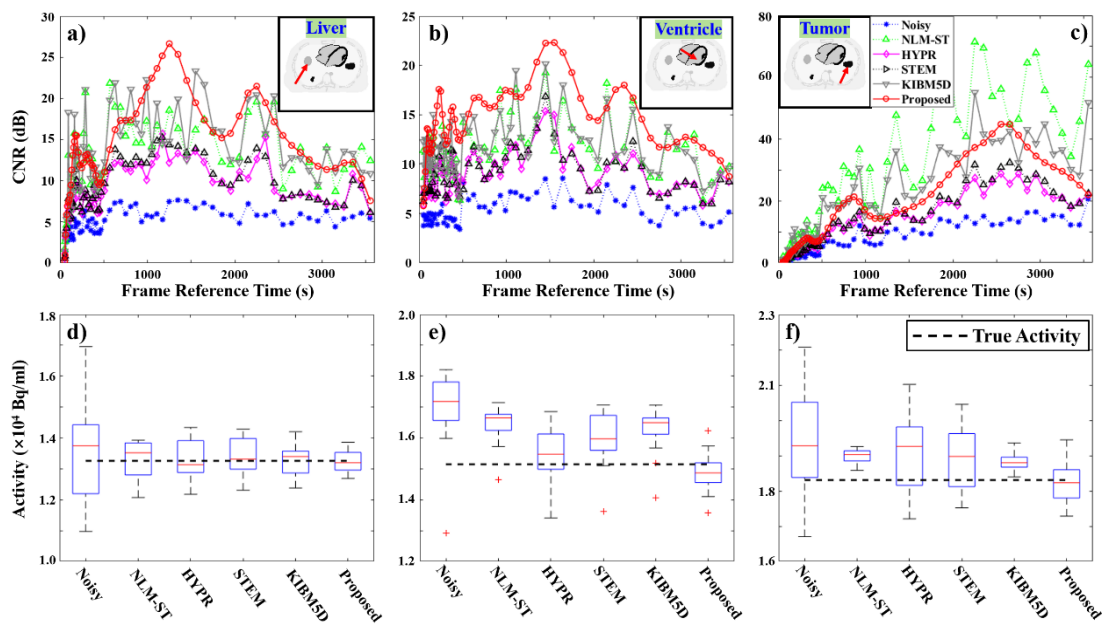

**Figure 5**. CNR and box plots of the simulative experiments. a)~c) Frame by frame CNR values of the ROIs of successively the liver, ventricle and tumor; d)~f) Box plot calculated based on ROIs extracted from the liver, ventricle and tumor in frame #59, #50 and #56, respectively.

### B. Clinical Image Experiments

Figure 6 shows the processing results of the clinical total-body dPET images in this study using different methods. A similar visual effect to the simulative study is that image noise is more effectively eliminated from the noisy HTR images by the Pitch-In method, and the results are even better than those of the corresponding LTR image. Tissue structure in the image is better restored by the proposed method, which is indicated by the clear aorta, as marked by the red arrows. For background tissue such as the lung, the noise is well controlled without visible sacrifice of spatial resolution (no remarkable blurring), as indicated by the blue arrows.

The calculated CNR values based on the regions of the aorta and tumor lesion are shown in Figure 7a)~b). For the aorta, the Pitch-In method gives the best contrast for most of the frames, and it yields the most stable contrast resolution of the tumor lesion over time compared to all the other methods. The box plots given in Figure 7c)~d) further illustrate the imaging accuracy and denoising effect of ROIs of the proposed Pitch-In method. Activity distributions of the aorta and tumor tissues are more reliable than those given by other methods refer to the boxes corresponding to the LTR images. Ki parametric images calculated from the proposed dynamic images with different methods are shown in Figure 8, from which we see that the denoising methods do not seriously skew the dynamic characteristics of the dynamic PET/CT image series except that the NLM-ST method brings some blurring effect to the resultant KI image.

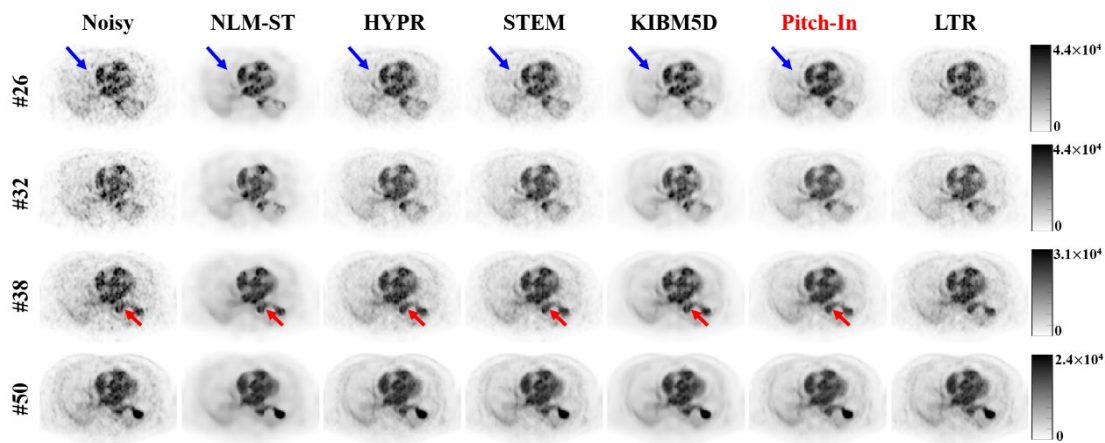

**Figure 6**. Results of the clinical image experiment. The unit of image intensity is Bq/ml. Number on the left indicates the frame index of the shown images in the HTR series, and the LTR images are the corresponding frames in the LTR series.

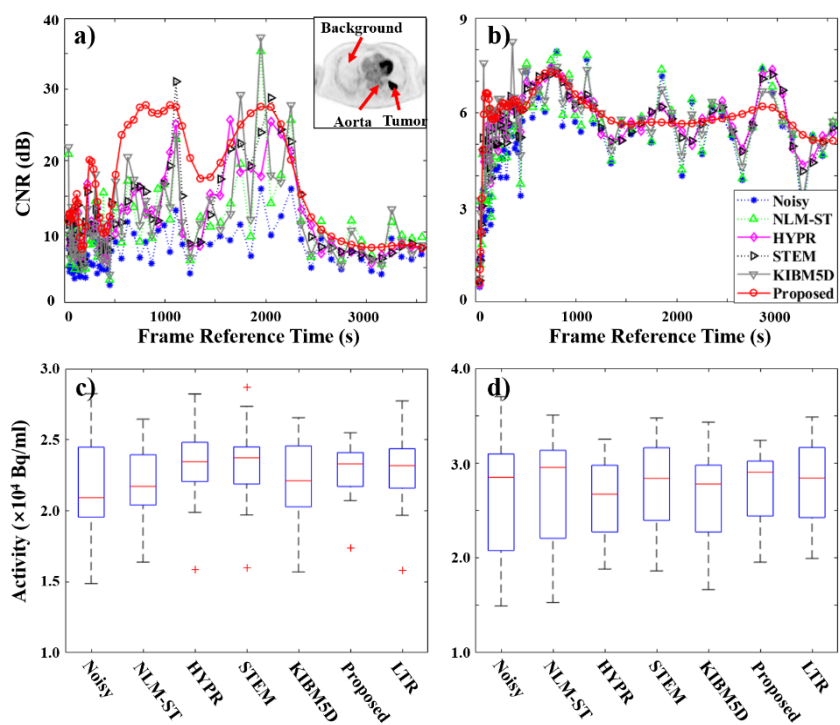

**Figure 7**. CNR and box plots of the clinical image experiments. a)~b) Frame by frame CNR values of the ROIs of successively the aorta and tumor; c)~d) Box plot calculated based on ROIs extracted from the aorta and tumor in frame #38 and # 47, respectively.

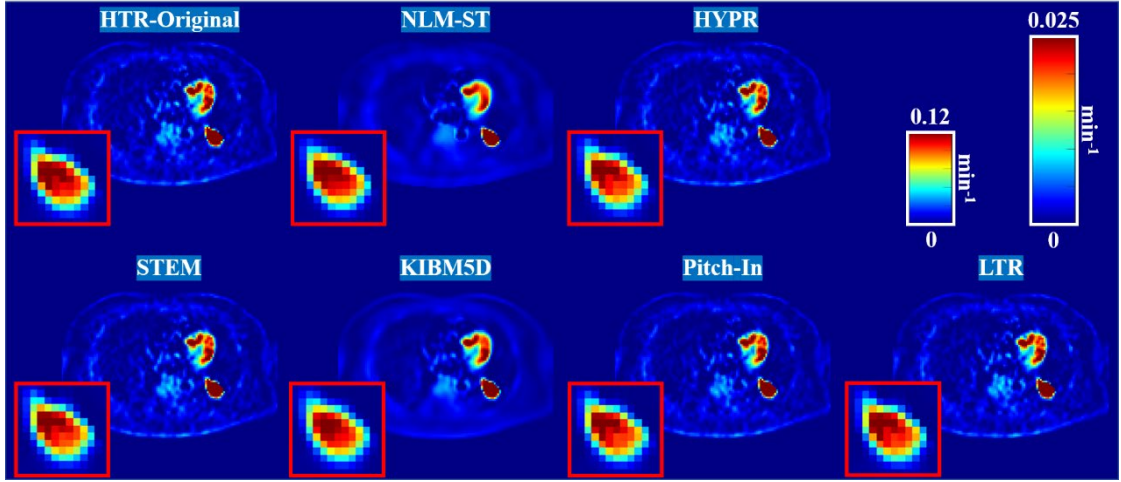

**Figure 8**. Clinical Ki parametric images calculated from original LTR and HTR dynamic images and the processed dynamic images with different methods. The subfigures are based on the tumor tissues.

## V. Discussion

In this study, we propose a method for high temporal resolution dPET image denoising based on the theory of third-order Hermite interpolation, which iteratively corrects every pixel-level TAC of the dynamic image data. The proposed method performs better in the task of denoising dPET images with shorter frame durations compared to several start-of-the-art methods and is proven to be an advanced and feasible technique for HTR dPET imaging. In this section, we give comments and discussions on the Pitch-In method.

### A. Advantages of the Pitch-in Method

One of the advantages of the Pitch-In method is that dPET images can be denoised on the premise of effective data fidelity. As described in Section II, the procedures of THI denoising and data fidelity based on the original noisy TAC are alternately carried out for processing dPET data, which promises that the pixel-level TACs are not distorted or oversmoothed by repeated THI operations. This is desirable since necessary spatial resolution of the images is required in clinical diagnosis. In addition, the denoising effect of Pitch-In is not sensitive to the intensity of the original image signal, as mentioned in Section IV, and therefore, spatial equilibrium in the processed images is guaranteed, namely, the background tissues with lower metabolism and the hypermetabolic tissues are denoised to nearly the same degree, avoiding unnatural or oversmoothed processing results.

Another reason that the proposed method does not blur the images is that the image data are processed entirely in the temporal domain, and no denoising operation in the spatial domain is involved. This is also the reason for the Pitch-In method to achieve advanced performance in terms of image structure restoration. As demonstrated in Section I, an advanced dPET imaging algorithm should never ignore the dynamic characteristics of the objective images, which is the underlying reason for the success of the proposed Pitch-In method: The direct application of dynamic information leads to such satisfying imaging performance. Figure 9 shows the visual effect of pixel-level TAC correction. As shown, the noisy TAC is processed to be smoother, while a temporal evolution tendency is not destroyed.

Since the proposed method can process the dPET images based on only the current imaging data, it does not face the trouble of model generalization, and extra experience needed for the effective implementation of the Pitch-In method includes only the parameter settings and the truncation of the very early noneffective frames. Therefore, it is expected to have higher feasibility for clinical application for HTR dPET imaging than many learning-based methods.

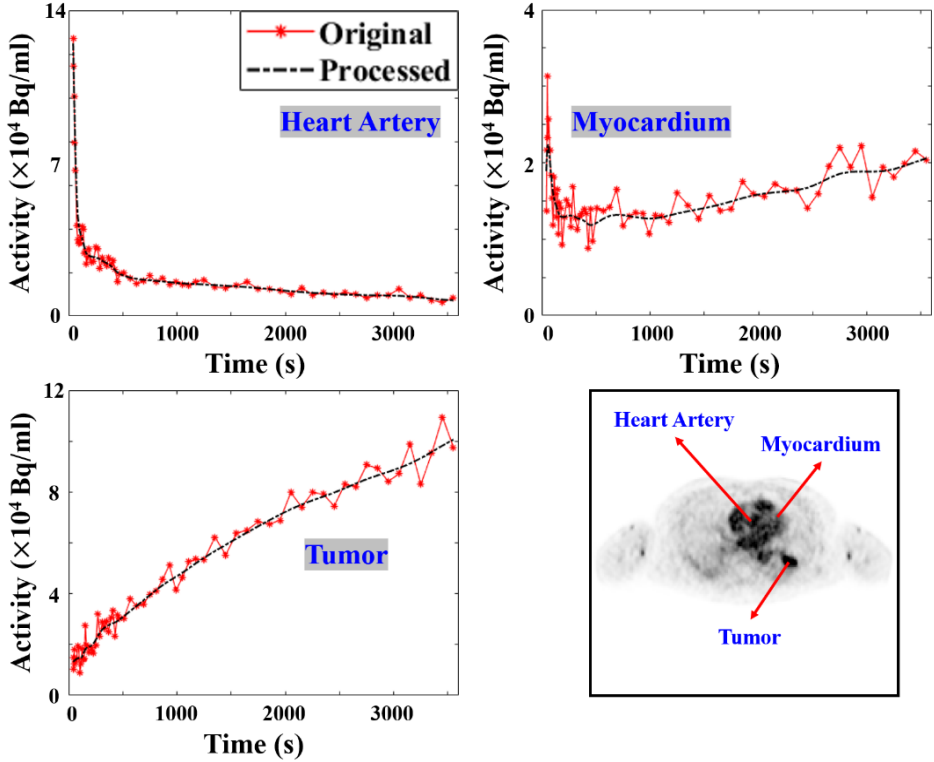

**Figure 9**. Processing effect of pixel-level TACs in the heart artery, myocardium and aorta tissues. Pixels are extracted from the regions indicated by the arrows.

## B. Parameter Selection

Several parameters need to be well set for the effective implementation of the proposed Pitch-In method. The settings used in the experiments in this work are given in Section II.E. Finely assigning all six listed parameters every time when dealing with a new dPET dataset seems inconvenient. The number of out-loops, $E$, the number of reference points for a single step of THI re-estimation, $n_r$, and the number of parallel re-estimations in one iteration of the THI subloop, $k_p$, can actually be fixed since they are not sensitive to data characteristics such as noise level, spatial structural characteristics of the dynamic images and the shape of the objective pixel-level TACs.

In contrast, the setting of the number of denoising iterations in one THI subloop, $E_t$,

and the THI updating relaxation parameter, $\eta$, play pivotal roles. Theoretically speaking, a larger $\eta$ should be accompanied by a smaller $E_t$, and vice versa. Therefore, in our study, we fixed $E_t$ to 30 and adjusted $\eta$. Figure 10a) shows the frame-by-frame PSNR values given by different $\eta$ with $E_t$=30 based on the simulation experiment, from which we see the trade-off between the quality of the very early frames and that of the later frames. An eclectic selection of $\eta$=0.6 was used for the optimization of the global imaging performance.

Another parameter that is worth discussing here is the overshoot parameter (suggested to be larger than 1) of the upper fidelity of TACs, $\beta_u$, which is designed for the protection of the peak of the processed TAC. The setting of $\beta_u$ should be based on the noise level of the original dPET data. Noisier data need a smaller $\beta_u$ since the peak of the curve in such data is more likely to be attributed to noise in the temporal domain. Figure 10b) shows the quantitative PSNR results given by different $\beta_u$ values tested. Based on the similar logic used for selecting $\eta$, $\beta_u$=0.6 was applied for the globally optimized imaging performance.

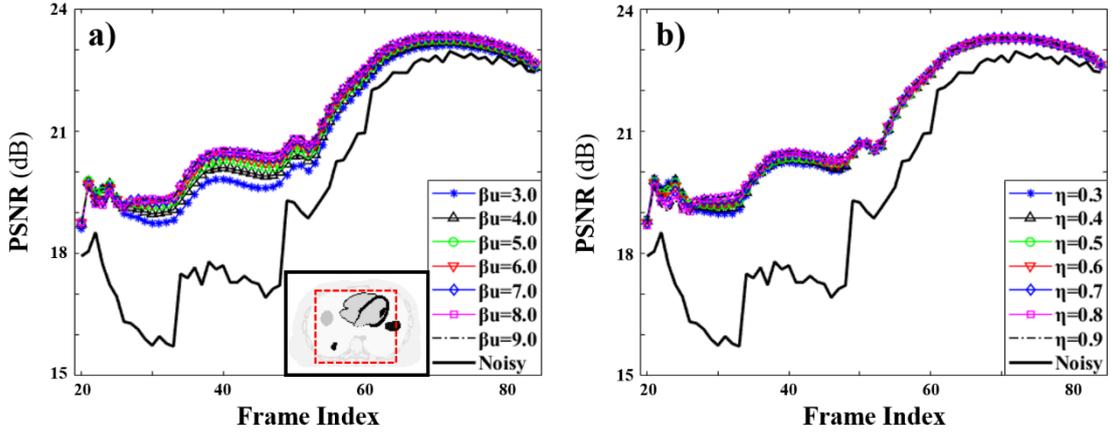

**Figure 10**. PSNR values yielded by different selection of $\beta_u$ and $\eta$. a) Frame-by-frame PSNR calculated based images processed with $\beta_u$ from 3.0 to 9.0 and $E_t$ fixed at 30; b) Frame-by-frame PSNR calculated based images processed with $\eta$ from 0.3 to 0.9 and $E_t$=30, $\eta$=0.6. The red dash box indicates the ROI used for PSNR calculation.

## C. Convergence of the Method

In this study, we designed a quantitative index, the mean absolute residual (MAR), to measure the convergence of the implementations of the Pitch-In method, which is calculated by

$$MAR^n = \frac{\sqrt{\sum_{j=1}^{N}\sum_{i=1}^{T}\left(c_{i,j}^n - c_{i,j}^{n-1}\right)^2}}{N \times T}, \tag{9}$$

where $n$ denotes the iteration index. $N$ and $T$ are the number of pixels in the applied image mask and the number of effective frames, respectively. Figure 11 shows the MAR curves of the simulative and clinical image experiments, from which we can see that 20 out-loop iterations are enough for the proposed Pitch-In method to reach convergence.

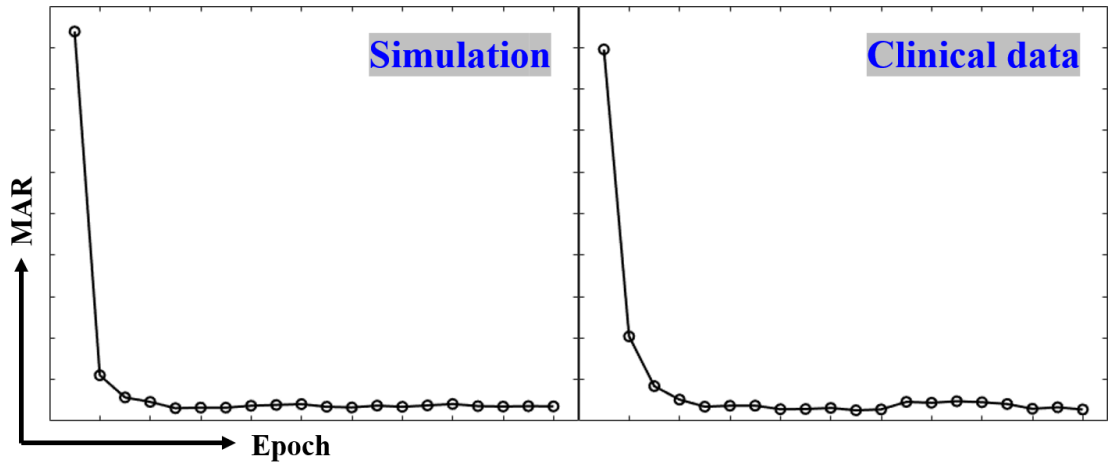

**Figure 11**. MAR curves of the simulative and clinical image experiments

### D. Potential defect and futural development of Pitch-In

The proposed Pitch-In method denoises dynamic PET TAC by applying curve interpolation in the temporal domain; therefore, if the operation parameters are not well adjusted, it may flatten the TACs with large net second-order derivatives, such as those for the aorta and liver, although empirical data fidelity is involved. Given this, to improve this algorithm, we may build an appropriate global cost function that depicts a combination of TAC noise and TAC distortion, based on which a more effective and stable iterative TAC correction method can be derived.

## VI. Conclusion

The Pitch-In method proposed in this work is effective for denoising dPET images with ultrashort frame durations. Compared to several state-of-the-art methods, the Pitch-In method has superior performance in terms of image noise reduction, imaging accuracy and imaging contrasting resolution. Moreover, the Pitch-In method has higher feasibility for clinical application compared to learning-based image processing methods or model-based image reconstruction methods.